\documentclass[11pt]{article}
\usepackage{moriond,epsfig}

\def\be{\begin{equation}}
\def\ee{\end{equation}}
\def\bea{\begin{eqnarray}}
\def\eea{\end{eqnarray}}

\begin{document}
\vspace*{2cm}
\title{Polarized CMB Foregrounds: What do we know and how bad is it?}

\author{Clive Dickinson }

\address{Jodrell Bank Centre for Astrophysics, School of Physics \& Astronomy, \\
University of Manchester, Oxford Rd, Manchester, M13 9PL, U.K.}

\maketitle\abstracts{Polarized foregrounds are going to be a serious challenge for detecting CMB cosmological B-modes. Both diffuse Galactic emission and extragalactic sources contribute significantly to the power spectrum on large angular scales. At low frequencies, Galactic synchrotron emission will dominate with fractional polarization $\sim 20-40\%$ at high latitudes while radio sources can contribute significantly even on large ($\sim 1^{\circ}$) angular scales. Nevertheless, simulations suggest that a detection at the level of $r=0.001$ might be achievable if the foregrounds are not too complex. }


\section{CMB foregrounds overview}

For high sensitivity measurements, once systematics are made negligible, component separation to remove foregrounds represents the ultimate limit to the precision in which the CMB, and therefore cosmological parameters, can be measured. Diffuse Galactic radiation consists of at least 3 components including synchrotron emission produced by relativistic electrons spiralling in the Galactic magnetic field, free-free emission from electrons accelerated by ionized bas and thermal dust emission due to black-body radiation from dust grains at temperatures of a few tens of degrees Kelvin. Other mechanisms are though to contribute at some level. In particular, electro-dipole emission from ultra-small rapidly spinning dust grains may be a significant contributor at frequencies $< 100$~GHz. Fortunately, at least for cosmologists, the CMB anisotropies in total-intensity are larger than the Galactic emission, over a significant fraction of the sky and over a few decades in frequency range i.e. $\sim 30-150$~GHz. Extragalactic sources, which are typically point-like relative to the experimental beam, are a major foreground on small angular scales, typically $<1^{\circ}$ or $\ell >200$. Their removal is usually achieved by masking/fitting the brightest sources, and making a statistical correction of the residual sources in the power spectrum.

The situation for polarization is different. Both Galactic and extragalactic radiation are significantly polarized, at the few to tens of a per cent level. Although the CMB E-mode fluctuations are at the 10\% level, the B-mode fluctuations are at least an order of magnitude lower than this, and possibly much smaller. It is therefore quite clear that CMB polarization measurements will be seriously affected by foregrounds. In this article, I summarise some new results for polarization of diffuse Galactic emission and extragalactic sources.

\begin{figure}
\begin{center}
\psfig{figure=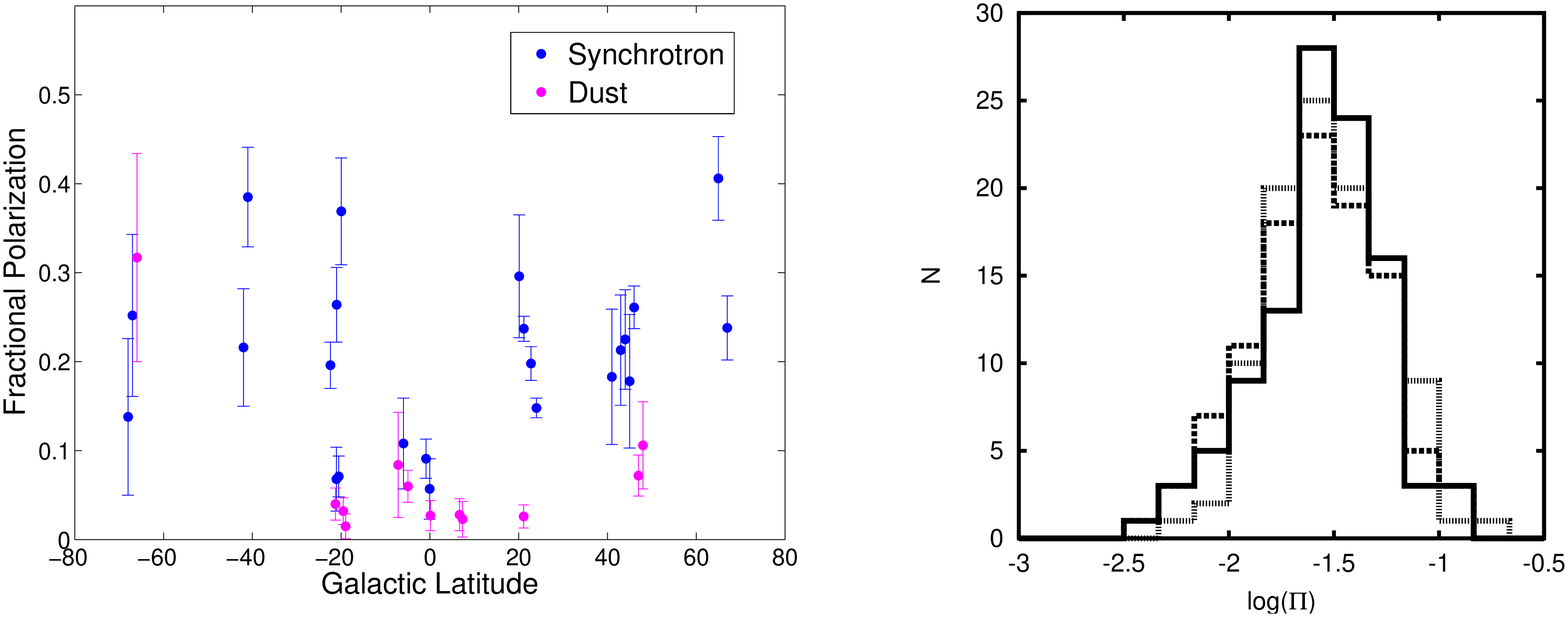,width=6.5in,angle=0}
\caption{{\it Left}: Fractional polarization of diffuse foregrounds as a function of Galactic latitude. {\it Right}: Fractional polarization of radio sources. \label{fig:fracpol}}
\end{center}
\end{figure}

\section{Template analysis of WMAP data in polarization}

Very little is known about the details of diffuse polarized foregrounds. Kogut et al.\cite{Kogut07} studied the global properties of the synchrotron and dust polarized emission, assuming the synchrotron polarization angle given by WMAP K-band and dust polarization angle given by a model map of starlight absorption. More recently, Macellari et al.\cite{Macellari10} used a template cross-correlation analysis by fitting total-intensity templates to the polarization data directly. The analysis amounts to fitting total-intensity maps to polarized intensity ($P=\sqrt{Q^2+U^2}$) at a resolution of $N_{side}=32$, by minimizing the $\chi^2$. A statistical noise bias was subtracted from $P$ and correlations between $Q$ and $U$ were taken into account, by using the WMAP noise covariance matrix supplied by the WMAP team, degraded to $N_{side}=32$. Pixel-pixel correlations were not taken into account, but are typically small ($\sim 1\%$). The main limitation of this method is that if the polarization angles of individual components are not the same, cross-terms appear that can bias the result. However, we do not expect the angles to be significantly different on large angular scales.

The left panel of Fig.~\ref{fig:fracpol} shows the fractional polarizations for the synchrotron- and dust- correlated components at K-band (23~GHz) as a function of Galactic latitude. The synchrotron polarization fraction is low ($\sim 5\%$) at low Galactic latitudes, as expected from depolarization along the line-of-sight. At high latitudes, the synchrotron fractional polarization increases to $\sim 15-40\%$; at $|b|>20^{\circ}$ the average is 19\%. We detect a dust-correlated signal, with an average polarization fraction of $2.9\pm0.6\%$. This is consistent with the expectation for spinning dust, although magneto-dipole emission cannot be ruled out. The H$\alpha$-correlated signal, expected to be due to free-free emission, has little or no polarization, with an all-sky average of $0.6\pm0.7\%$.

\section{Contribution of polarized extragalactic sources}

\begin{figure}
\begin{center}
\psfig{figure=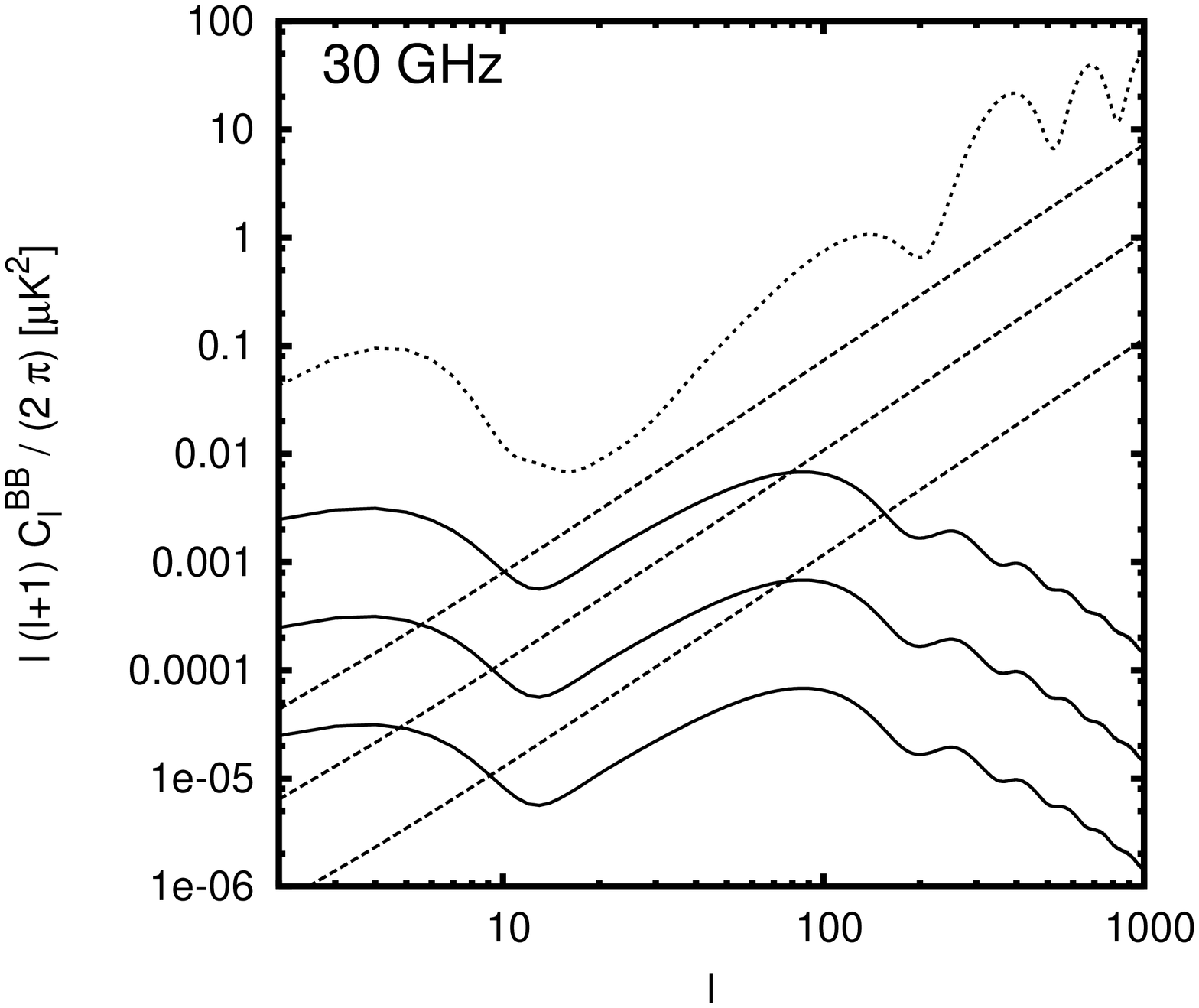,width=2in,angle=-90}
\psfig{figure=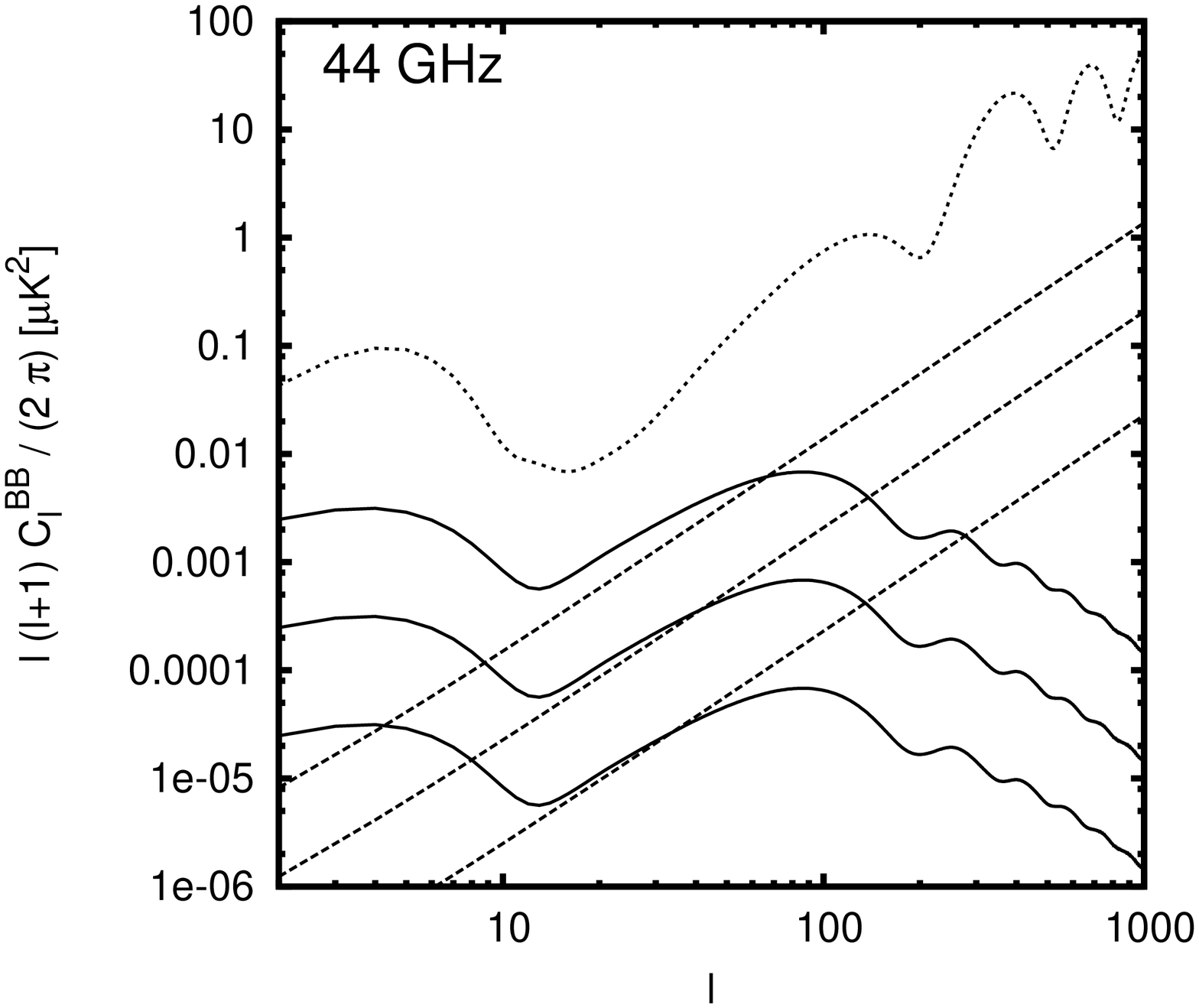,width=2in,angle=-90}
\psfig{figure=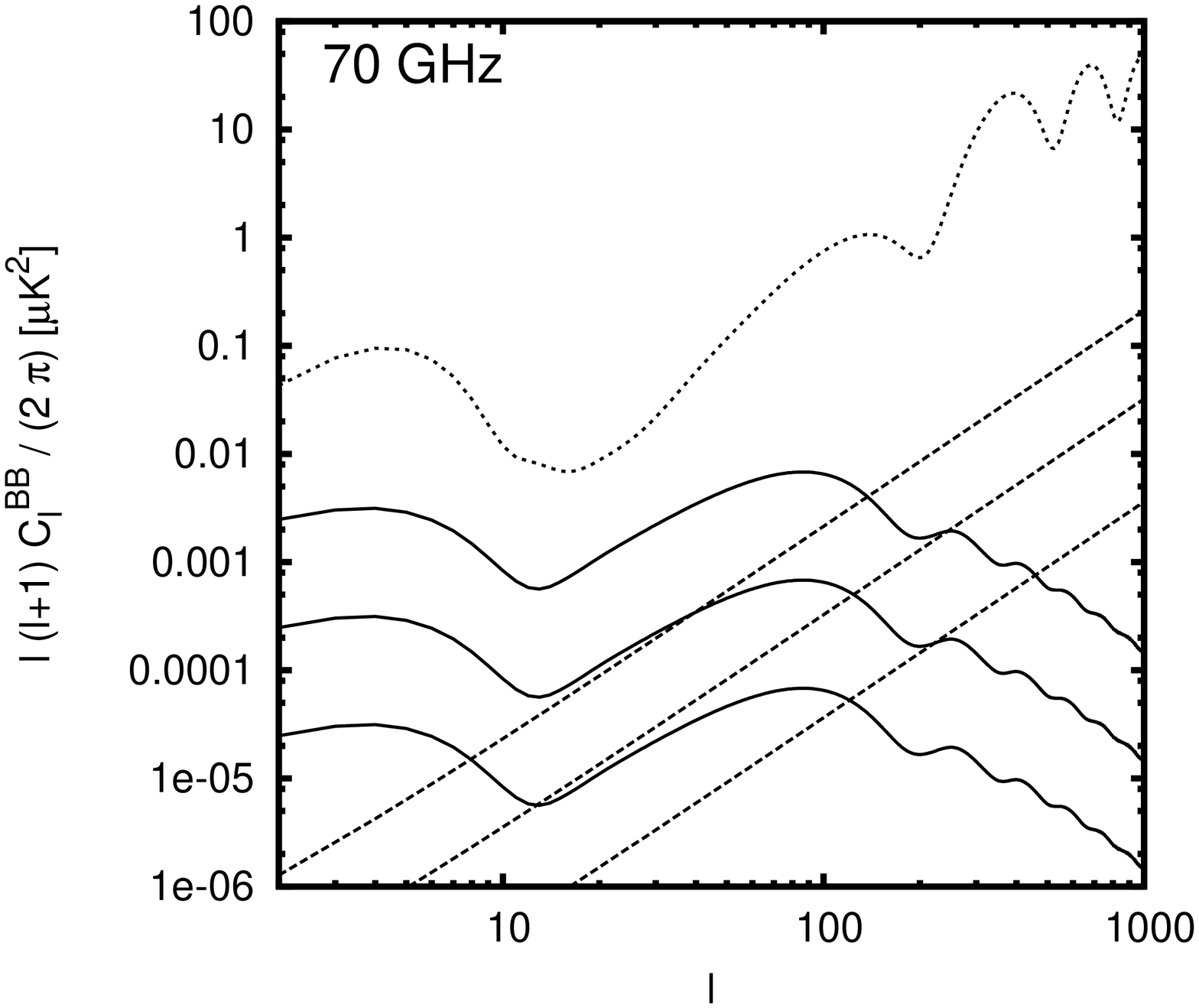,width=2in,angle=-90}
\psfig{figure=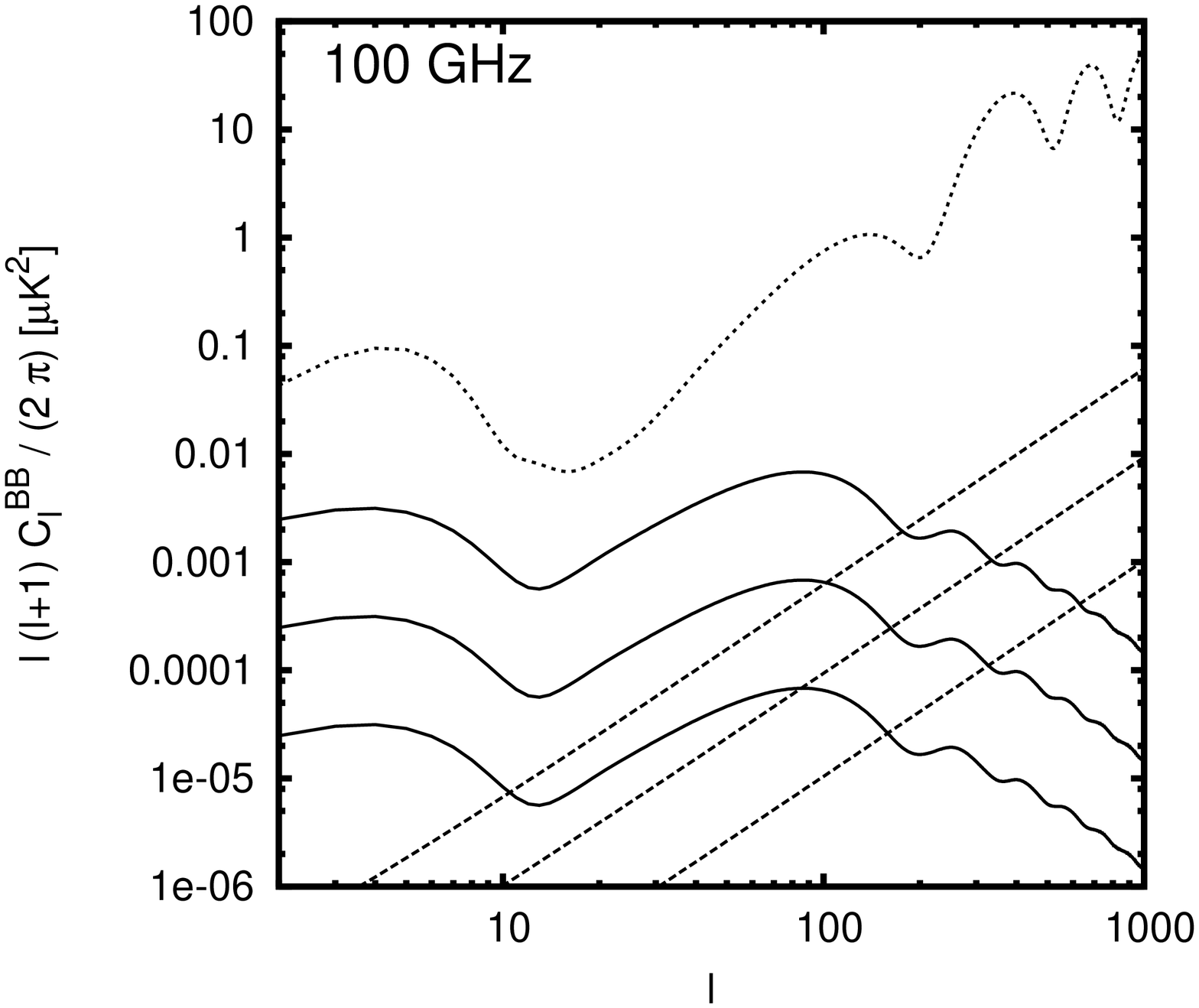,width=2in,angle=-90}
\caption{Projected power spectra of radio sources at 30, 44, 70 and 100~GHz. The dotted line shows the theoretical CMB E-mode spectrum while the solid lines are the CMB B-mode spectra for $r=0.1, 0.01, 0.001$. The dashed lines show the power spectra of radio sources for different intensity flux density cut-offs, $S_{\rm cut}=1, 0.1, 0.01$~Jy. \label{fig:sources_ps}}
\end{center}
\end{figure}

Extragalactic sources are known to exhibit polarization. At frequencies $>100$~GHz, there is very little information at all, except to note that the polarization must be relatively small for most galaxies (e.g. Seiffert et al.\cite{Seiffert07}). At frequencies $<100$~GHz, radio surveys such as the NVSS at 1.4~GHz (Condon et al.\cite{Condon98}) have accurately characterised source counts down to a few mJy. However, the polarization properties are still not well known, except for the brightest sources. Recently, Jackson et al.\cite{Jackson10} observed bright ($>1$~Jy) sources detected by WMAP (Hinshaw et al.\cite{Hinshaw09}) with the Very Large Array (VLA) in polarization at 8.4, 22 and 43~GHz. The right panel of Fig.~\ref{fig:fracpol} shows the distribution of polarization fractions, $\Pi$, for sources detected at all 3 VLA frequencies. The distributions at 8.4, 22 and 43~GHz are almost identical and can be approximated by a Gaussian distribution in log$(\Pi)$. The median value is $\simeq 2\%$ with an average of $\simeq 3.5\%$. This allows us to estimate the contribution of point sources to the CMB power spectrum in polarization. Fig.~\ref{fig:sources_ps} shows the source power spectra at 4 frequencies and for 3 different flux cut-off values. We have assumed no clustering of radio sources, and that the spectra and polarization fractions do not vary at lower flux densities. It is clear that, even at large angular scales ($\sim 1^{\circ}$), extragalactic sources will need to be removed if one is to try to detect $r\sim 0.001$. In particular, at low frequencies the power from radio sources will dominate B-modes at $r=0.001$ even when a large number of sources have been removed. To measure CMB B-modes at this level, accurate statistical corrections in the power spectrum will need to be applied.

\section{What tensor-to-scalar ratio can be achieved?}

Without detailed knowledge of polarized foregrounds, it is difficult to calculate the lowest value of the tensor-to-scalar ratio, $r$, that might be achieved with a future CMB mission (e.g. CMBpol/EPIC, Bpol). However, using simple models, normalized to the approximate levels that we see in current experiments, we can estimate the ultimate $r$-value, assuming the foregrounds are relatively simple. It should be remembered that component separation is likely to set the limit on the lowest $r$-value that can be achieved.

As part of a white paper in preparation for the U.S. 2010 decadal review and the CMBpol design study for a future CMB polarisation satellite, Dunkley et al.\cite{Dunkley09} investigated the issue of foreground removal. Using the Planck Sky Model, a simulation of diffuse foregrounds was produced at a range of frequencies, as proposed for a particular configuration of the EPIC satellite. In polarization, this consisted of synchrotron and thermal dust components, based on the model of Miville-Desch{\^e}nes et al.\cite{Miville-Deschenes08}. White noise was added, in accordance to the EPIC design. No extragalactic sources were added and CMB lensing was not considered.

A number of techniques were used to estimate the sensitivity to $r$ that could be achieved, including a Fisher-matrix calculation, parametric fitting with various assumptions (Eriksen et al.\cite{Eriksen08}), and a blind component separation method (Delabrouille et al.\cite{Delabrouille03}). Two $\ell$ ranges were considered, to take into account that lensing of E-modes into B-modes will be a major challenge for $\ell >15$. The results are summarised in Table~\ref{tab:rvalues}. The results suggest that $r\sim 0.01$ should be relatively easily achievable, so long as the foregrounds are not significantly more complex than we expect. If we assume that lensing B-modes can be removed, then $r<0.001$ may be attainable with ultra-sensitive instruments; see also Betoule et al.\cite{Betoule09}. However, foregrounds (including lensing) are likely to make detecting $r<10^{-4}$ impossible. It is clear, however, that accurate data over the frequency range of a few GHz to $\sim 1$~THz is needed to characterize and understand foregrounds so that they can be accurately removed from sensitive CMB data.

\begin{table}[t]
\caption{Forecasted $1\sigma$ uncertainties on the tensor-to-scalar ratio, $r$, for a fiducial value $r=0.01$.\label{tab:rvalues}}
\vspace{0.4cm}
\begin{center}
\begin{tabular}{c|c|c|c}
Method &Description &$\ell < 15$ &$\ell < 150$ \\
\hline
Fisher &Assumed $10\%$ residual foregrounds &0.014 &0.00052  \\ 
Parametric &Power-law indices  &0.003   &--   \\
Blind      &SMICA   &--       &0.00055   \\
\end{tabular}
\end{center}
\end{table}

\section*{Acknowledgments}
CD acknowldeges an STFC Advanced Fellowship and an ERC grant under the FP7.


\section*{References}
\begin{thebibliography}{99}

\bibitem{Battye10} Battye, R.~A., Browne, I.~W.~A., Peel, M.W., Jackson, N.~J., Dickinson, C.\ 2009, MNRAS, submitted (arXiv:1003.5846).

\bibitem{Betoule09} Betoule, M., Pierpaoli, E., Delabrouille, J., Le Jeune, M., \& Cardoso, J.-F.\ 2009, A\&A, 503, 691

\bibitem{Condon98} Condon, J.~J., Cotton, W.~D., Greisen, E.~W., Yin, Q.~F., Perley, R.~A., Taylor, G.~B., \& Broderick, J.~J.\ 1998, AJ, 115, 1693. 

\bibitem{Delabrouille03} Delabrouille, J., Cardoso, J.-F., \& Patanchon, G.\ 2003, MNRAS, 346, 1089

\bibitem{Dunkley09} Dunkley, J., et al.\ 2009, American Institute of Physics Conference Series, 1141, 222

\bibitem{Eriksen08} Eriksen, H.~K., Jewell, J.~B., Dickinson, C., Banday, A.~J., G{\'o}rski, K.~M.,\& Lawrence, C.~R.\ 2008, ApJ, 676, 10

\bibitem{Hinshaw09} Hinshaw, G., et al.\ 2009, ApJS, 180, 225.

\bibitem{Jackson10} Jackson, N., Browne, I.~W.~A., Battye, R.~A., Gabuzda, D., \& Taylor, A.~C.\ 2010, MNRAS, 401, 1388.

\bibitem{Kogut07} Kogut, A., et al.\ 2007, ApJ, 665, 355

\bibitem{Macellari10} Macellari, N., Piepaoli, E., Dickinson, C., Vaillancourt, J..\ 2010, MNRAS, submitted.

\bibitem{Miville-Deschenes08} Miville-Desch{\^e}nes, M.-A., Ysard, N., Lavabre, A., Ponthieu, N., Mac{\'{\i}}as-P{\'e}rez, J.~F., Aumont, J., \& Bernard, J.~P.\ 2008, A\&A, 490, 1093

\bibitem{Seiffert07} Seiffert, M., Borys, C., Scott, D., \& Halpern, M.\ 2007, MNRAS, 374, 409

\end{thebibliography}
\end{document}